\documentclass[aps,noshowpacs,nofootinbib,superscriptaddress,notitlepage,preprintnumbers,amsmath,amssymb]{revtex4-1}
\usepackage{multirow}
\usepackage{graphicx}
\usepackage{hyperref}
\usepackage{microtype}

\newcommand{\RNum}[1]{\uppercase\expandafter{\romannumeral #1\relax}}

\def\be{\begin{equation}}
\def\ee{\end{equation}}
\def\bea{\begin{eqnarray}}
\def\eea{\end{eqnarray}}
\def\bear{\begin{array}}
\def\ear{\end{array}}
\def\bfig{\begin{figure}}
\def\efig{\end{figure}}
\def\bcen{\begin{center}}
\def\ecen{\end{center}}
\def\bi{\begin{itemize}}
\def\ei{\end{itemize}}

\def\raw{\rightarrow}

\def\slashchar#1{\setbox0=\hbox{$#1$}
   \dimen0=\wd0 \setbox1=\hbox{/} \dimen1=\wd1
   \ifdim\dimen0>\dimen1 \rlap{\hbox to \dimen0{\hfil/\hfil}} #1
   \else  \rlap{\hbox to \dimen1{\hfil$#1$\hfil}} / \fi}

\begin{document}
\title {Photon emission in neutral current interactions at the T2K experiment}

\author{E. Wang}
\affiliation{Department of Physics, Zhengzhou University, Zhengzhou, Henan 450001, China}

\author{L. Alvarez-Ruso}
\affiliation{Instituto de F\'\i sica Corpuscular (IFIC), Centro
Mixto CSIC-Universidad de Valencia, Institutos de Investigaci\'on de
Paterna, E-46071 Valencia, Spain}

\author{Y. Hayato}
\affiliation{University of Tokyo, Institute for Cosmic Ray Research, Kamioka Observatory, Kamioka, Japan}

\author{K. Mahn}
\affiliation{Michigan State University, Department of Physics and Astronomy, 
East Lansing, Michigan, U.S.A.}

\author{J. Nieves}
\affiliation{Instituto de F\'\i sica Corpuscular (IFIC), Centro
Mixto CSIC-Universidad de Valencia, Institutos de Investigaci\'on de
Paterna, E-46071 Valencia, Spain}


\begin{abstract}
We have applied a microscopic model for single photon emission in neutral current interactions on nucleons and nuclei to determine the number and distributions of such events at the Super-Kamiokande detector, for the flux and beam exposure of the T2K experiment in neutrino mode. These reactions represent an irreducible background in electron-(anti)neutrino appearance measurements aimed at a precise measurement of mixing angle $\theta_{13}$ and the $CP$ violating phase. We have obtained a total number of photon events that is twice larger than the one from the NEUT event generator (version 5.1.4.2) used in the analysis of T2K data. Detailed comparisons of energy and angular distributions for the $\nu_\mu$ and $\bar\nu_\mu$ fluxes have also been performed.  
\end{abstract}

\maketitle
 
\section{Introduction}
\label{sec:introduction}
More than 15 years of dedicated experimental studies have established the oscillations of three flavors of massive neutrinos. Recently, a new piece in the puzzle has been added by the determination of the so-called reactor neutrino mixing angle $\theta_{13}$. The first indication with a $2.5 \,\sigma$ significance of a nonzero value of $\theta_{13}$ was provided by the T2K experiment in a study of $\nu_e$ appearance in a $\nu_\mu$ beam~\cite{Abe:2011sj}. Afterwards it has been precisely measured from $\bar\nu_e$ disappearance in nuclear reactor neutrino experiments~\cite{An:2013zwz,Ahn:2012nd,Abe:2011fz}. The significance of the T2K $\nu_e$ appearance result has now reached $7.3 \, \sigma$~\cite{Abe:2013hdq}. The increasing precision in these experiments creates a window of opportunity to determine the $CP$ violating phase in the lepton sector. Indeed, the tension between reactor data and T2K favors a $\delta_{CP} = -\pi/2$~\cite{Abe:2013hdq} at 90\%~C.L., although the picture is still far from clear because the MINOS combined $\nu_\mu$ disappearance and $\nu_e$ appearance prefers a $\delta_{CP} = \pi/2$~\cite{Adamson:2014vgd}.

Further progress in this direction requires a better control over systematic errors and, in particular, of irreducible backgrounds. For this purpose, a better understanding of neutrino interactions with matter is mandatory. The ongoing effort in this direction encompasses more precise measurements of different (anti)neutrino cross sections on nuclear targets, theoretical work aimed at a better description of weak reactions on both nucleons and nuclei, and improvement of the Monte Carlo simulation codes; see the reviews of Refs.~\cite{Morfin:2012kn,Formaggio:2013kya,Alvarez-Ruso:2014bla} for different aspects of these problems.    

Super-Kamiokande (SK), the far detector of the T2K experiment, is a water Cherenkov detector and, as such, is incapable of discriminating the diffuse rings of $e^{\pm}$ originated in charged current interactions by electron neutrinos from those created by photons. The largest part of such a background originates in $\pi^0$ production in neutral current (NC) interactions (NC$\pi^0$)  when the two photons from $\pi^0 \raw \gamma \gamma$ produce overlapping rings or when one of them is not observed. Another relevant source is the NC single photon emission (NC$\gamma$) reaction. Although NC$\pi^0$ has a larger cross section than NC$\gamma$, the $\pi^0$ background can be reduced with dedicated reconstruction algorithms, while the NC$\gamma$ one remains irreducible. Indeed, in the latest T2K analysis~\cite{Abe:2013hdq}, the NC$\pi^0$ background was reduced by 69\% with respect to the previous appearance selection~\cite{Abe:2013xua}. In this context, the relative relevance of the NC$\gamma$ channel is significantly enhanced.   

The interest in a detailed theoretical study of the NC$\gamma$ reaction~\cite{Hill:2009ek,Serot:2012rd,Zhang:2012aka,Zhang:2012xi,Wang:2013wva} followed the observation of an excess of electron-like events at low reconstructed energies in the MiniBooNE detector, in both neutrino and antineutrino modes~\cite{AguilarArevalo:2008rc,Aguilar-Arevalo:2013pmq}. It was suggested that an anomalous contribution to NC photon emission could be responsible for this~\cite{Harvey:2007rd}. However,  in spite of the fact that first studies indicated that NC$\gamma$ indeed accounted for the excess~\cite{Hill:2010zy}, more recent analyses, considering nuclear effects and realistic acceptance corrections~\cite{Zhang:2012xn,Wang:2014nat}, obtain a number of photon-induced electron-like events which is consistent with the estimate made by MiniBooNE, using a poor resonance production model tuned to the experiment's own NC$\pi^0$ measurement. It has also been proposed that additional photons from electromagnetic heavy neutrino decays could be at the heart of the MiniBooNE anomaly~\cite{Gninenko:2009ks,Masip:2012ke}, which would have implications for other experiments such as T2K and MicroBooNE. 

Here, we apply the microscopic model of Ref.~\cite{Wang:2013wva}, briefly described in Sec.~\ref{sec:model},
to predict the number of NC$\gamma$ events at the SK detector, as well as their energy and angular distributions, for the flux and beam exposure of the latest T2K $\nu_e$ appearance study~\cite{Abe:2013hdq}. The event number and distributions are compared to those of the NEUT event generator~\cite{Hayato:2009zz}, from which the T2K estimates are obtained. 

\section{Outline of the theoretical model}
\label{sec:model}

The main features of our model for NC photon emission induced by (anti)neutrinos on nucleons and nuclei are presented in the following. A detailed account is given in Ref.~\cite{Wang:2013wva}, with comparisons to previous results. A concise overview of the different models can be found in Ref.~\cite{Alvarez-Ruso:2014bla}.

\subsection{NC$\gamma$ on nucleons}
\label{subsec:theory_nucleon}

The differential cross section for the reactions
\begin{eqnarray}
 \nu_l (k) +\, N(p)  &\to & \nu_l (k^\prime) + N(p^\prime) +\gamma(k_{\gamma}) \, , \nonumber \\
 \bar \nu_l (k) +\, N(p)  &\to & \bar \nu_l (k^\prime) + N(p^\prime) +\gamma(k_{\gamma}) \,  \nonumber
\label{eq:reaction_nucleon}
\end{eqnarray}
on nucleons (protons or neutrons) in the laboratory frame is given by
\begin{equation}
\frac{d^{\,3}\sigma_{(\nu,\bar\nu)}}{dE_\gamma d\Omega(\hat{k}_\gamma)} =
    \frac{E_\gamma}{ |\vec{k}|}\frac{G^2}{16\pi^2}
     \int  \frac{d^3k'}{|\vec{k}^{\prime}\,|}
    L^{(\nu,\bar\nu)}_{\mu\sigma}W^{\mu\sigma} 
\label{eq:sec} \, ,
\end{equation}
where $k=(E_\nu,\vec{k})$, $k'=(E',\vec{k}')$ and $k_\gamma = (E_\gamma, \vec{k}_\gamma)$; $G$ denotes the Fermi constant. The leptonic tensor 
\be
L_{\mu\sigma}^{(\nu,\bar\nu)} =
 k^\prime_\mu k_\sigma +k^\prime_\sigma k_\mu
+ g_{\mu\sigma} \frac{q^2}2 \pm i
\epsilon_{\mu\sigma\alpha\beta}k^{\prime\alpha}k^\beta \,,
 \label{eq:lep}
\ee
with $q=k-k'$ and the fully antisymmetric tensor defined such that $\epsilon_{0123}= +1$, is contracted with the hadronic one 
\be
W^{\mu\sigma} = \frac{1}{4 m_N}\overline{\sum_{\rm
 spins}} \int\frac{d^3p^\prime}{(2\pi)^3} \frac{1}{2E^\prime_N}
  \delta^4(p^\prime+k_\gamma-q-p) \langle N \gamma |
 j^\mu(0) | N \rangle \langle N \gamma | j^\sigma(0) | N
 \rangle^* 
\label{eq:wmunu-nucleon} \,,
\ee
where $p=(m_N,\vec{0})$ and $p'=(E_N',\vec{p}\,')$. A sum (average) is performed over the final (initial) spin states. The matrix element of the hadronic current 
\begin{equation}
\langle N \gamma |
 j^\mu(0) | N \rangle = \bar u(p') \Gamma^{\mu \rho}_N u (p) \epsilon^*_\rho(k_\gamma) \,,
 \label{eq:amputated_amp}
\end{equation}
with $\epsilon(k_\gamma)$ being the photon polarization four vector, is defined~\cite{Wang:2013wva} by the set Feynman diagrams shown in Fig.~\ref{fig:Feynman}.
\bfig[h!]
\includegraphics[width=0.3\textwidth]{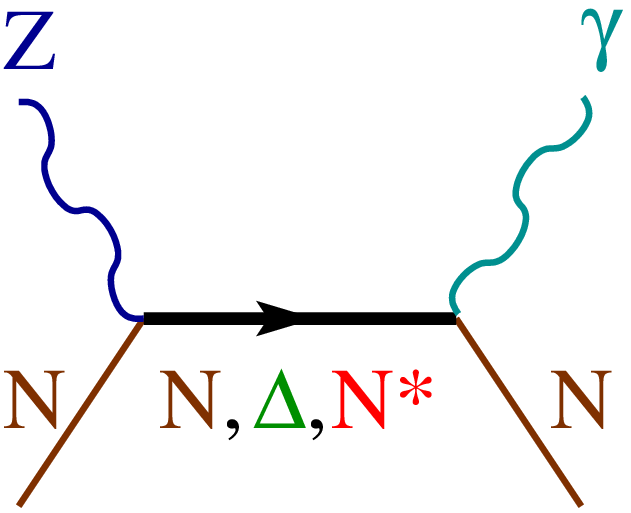}
\hspace{.05\textwidth} 
\includegraphics[width=0.3\textwidth]{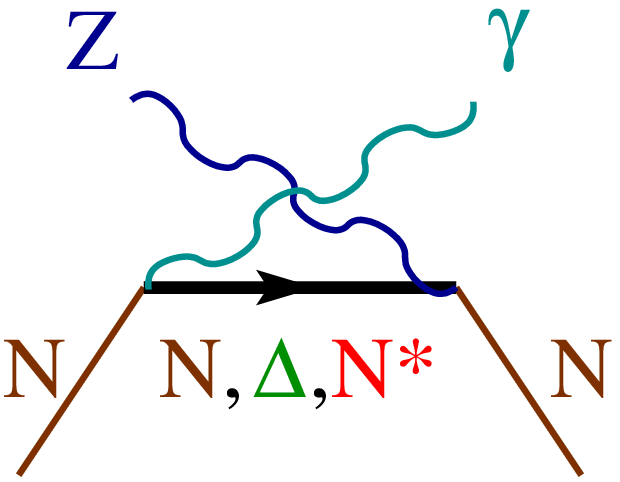}
\hspace{.05\textwidth} 
\includegraphics[width=0.25\textwidth]{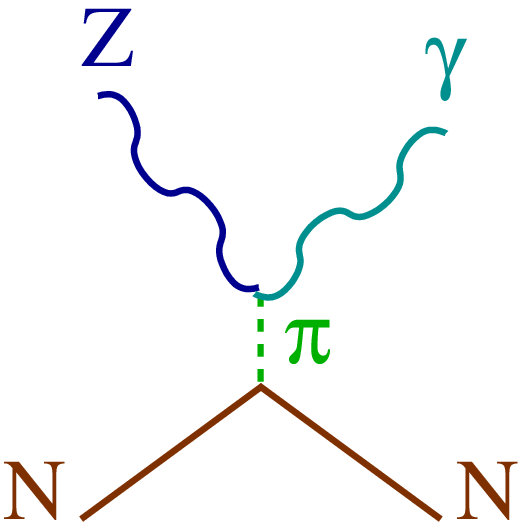}
\caption{\label{fig:Feynman}  Feynman diagrams for the hadronic current of NC photon emission off nucleons considered in Ref.~\cite{Wang:2013wva}. The first two diagrams represent direct and crossed baryon-pole terms with nucleons and nucleon resonances in the intermediate state: $BP$ and $CBP$ with $B=N$, $\Delta(1232)$, $N(1440)$, $N(1520)$, $N(1535)$. The third diagram denotes $t$-channel pion exchange: $\pi Ex$.}
\efig

At neutrino energies around and below 1~GeV, where most of the T2K flux is concentrated (see Fig.~\ref{fig:flux}), the NC$\gamma$  processes are dominated by the weak excitation of the $\Delta(1232)$ resonance followed by its radiative decay. Non-resonant nucleon-pole contributions that close to threshold, are fully constrained by the chiral symmetry are also important. We have also taken into account mechanisms with nucleon excitations from the second resonance region [$P_{11}$ $N(1440)$, $D_{13}$ $N(1520)$ and $S_{11}$ $N(1535)$] as intermediate states. Among them, the $N(1520)$ contributes most significantly (see Fig 5. of Ref.~\cite{Wang:2013wva}). The pion-exchange mechanism arises from the anomalous $Z\gamma\pi^0$ vertex and gives a very small contribution to the cross section~\cite{Wang:2013wva}. Other t-channel $\rho$ and $\omega$ exchange amplitudes, not considered in Ref.~\cite{Wang:2013wva}, coming from the corresponding anomalous vertices~\cite{Harvey:2007rd,Harada:2011xx} have turned out to be small~\cite{Hill:2009ek}, particularly according to the latest estimate~\cite{Rosner:2015fwa}.  

In terms of the amputated amplitudes introduced in Eq.~(\ref{eq:amputated_amp})
\be
W^{\mu\sigma}  = - \frac{1}{8 m_N} \int\frac{d^3p^\prime}{(2\pi)^3} \frac{1}{2E^\prime_N}
  \delta^4(p^\prime+k_\gamma-q-p) {\rm Tr}\left[
  (\slashchar{p}'+ m_N)\Gamma^{\mu\rho}(\slashchar{p}+ m_N) \gamma^0
  (\Gamma^\sigma_{.\,\rho})^\dagger \gamma^0 \right] \,,
\label{eq:wmunu-nucleon-avg}
\ee
with
\be
\Gamma^{\mu\rho}_N=\sum_a \Gamma^{\mu\rho}_a, \,\,
a=BP, CBP, \pi Ex,~\mathrm{and}~B=N, \Delta(1232), N(1440), N(1520), N(1535) \,.
\label{eq:hadronNR}
\ee
Explicit expressions for these amplitudes can be found in Ref.~\cite{Wang:2013wva}. They are functions of phenomenological nucleon and nucleon-to-resonance vector and axial form factors (partially) constrained by (quasi)elastic and pion production reactions induced by electrons and neutrinos on hydrogen and deuterium targets.

\subsection{NC photon emission in nuclei}
\label{subsec:nuclei}

On nuclear targets, NC photon emission can be incoherent, 
\begin{eqnarray}
  \nu_l (k) +\, A_Z  & \to& \nu_l (k^\prime) + \,
   \gamma(k_\gamma) +\, X \,,\nonumber \\  
    \bar\nu_l (k) +\, A_Z   &\to & \bar\nu_l (k^\prime) + \,
   \gamma(k_\gamma) +\, X\,, \nonumber
\label{eq:reacincl}
\end{eqnarray}
with the final nucleus in any excited state or broken, or coherent 
\begin{eqnarray}
  \nu_l (k) +\, A_Z|_{gs}(p_A)  &\to &\nu_l (k^\prime) + \, 
  A_Z|_{gs}(p^\prime_A) +\, \gamma(k_\gamma) \,, \nonumber \\   
\bar\nu_l (k) +\, A_Z|_{gs}(p_A) & \to &\bar\nu_l (k^\prime) + \, 
  A_Z|_{gs}(p^\prime_A) +\, \gamma(k_\gamma) \,, \nonumber
\label{eq:reac}
\end{eqnarray}
when the nucleus is left in the ground state. Equation~(\ref{eq:sec}) remains valid in both cases, with the hadronic tensor replaced by the pertinent one,  $W^{\mu\sigma}_{\mathrm{incoh}}$ or  $W^{\mu\sigma}_{\mathrm{coh}}$.  

\begin{figure}[h!]
\includegraphics[width=.7\textwidth]{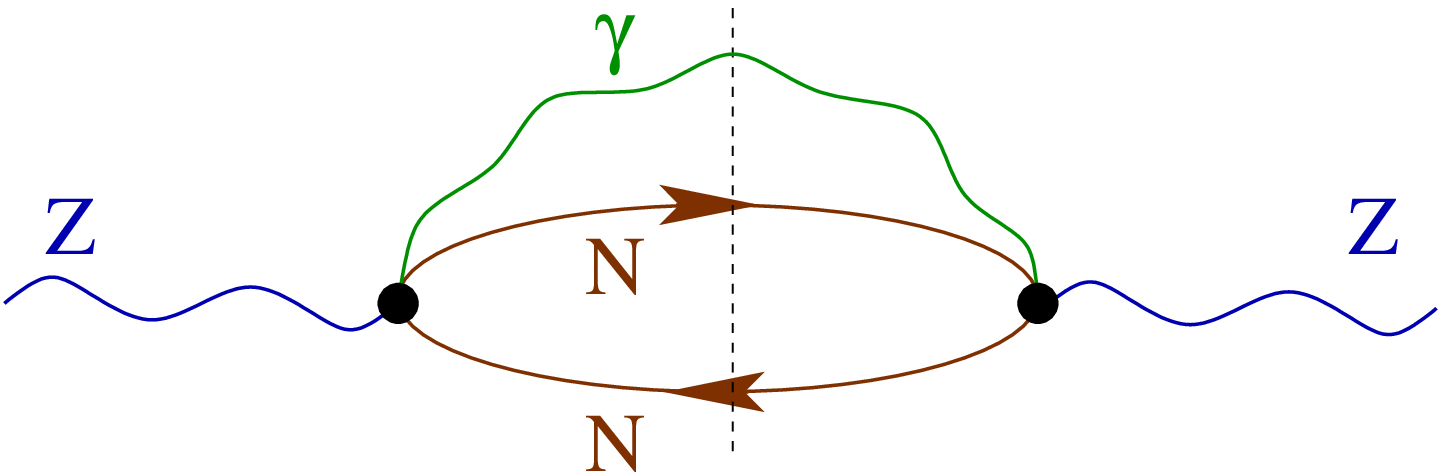}
\caption{Diagrammatic representation of the one-particle-one-hole-photon (1p1h$\gamma$) 
contributions to the $Z^0$ self-energy in nuclear matter. The black dots represent 
the $Z^0N \to \gamma N$ elementary amplitudes of Fig~\ref{fig:Feynman}. To obtain the imaginary part, the intermediate states intersected by the dashed line have to be placed on the mass shell.}\label{fig:1p1h}
\end{figure}
In the incoherent case, Ref.~\cite{Wang:2013wva} applies a many-body scheme for neutrino propagation in nuclear matter adapted to (semi)inclusive reactions on finite nuclei using the local density approximation~\cite{Nieves:2004wx}. In this framework, the hadronic tensor is obtained as the imaginary part of the contributions to the $Z$ selfenergy with a single photon in the intermediate state. To the lowest order in a density expansion, corresponding to a one-particle-one-hole nuclear excitation (Figure~\ref{fig:1p1h}) this tensor reads
\begin{equation}
W^{\mu\sigma}_{\mathrm{incoh}} \approx W^{\mu\nu}_{1{\rm p1h}\gamma} = \Theta(q^0) \Theta(q^0-E_\gamma) \frac{1}{(2 m_N)^2}  \sum_{N=p,n}  \int
\frac{d^3r }{(2\pi)^4} {\rm Im}\overline {U}_R(q-k_\gamma,k_F^N,k_F^{N}) A^{\nu\mu}_{N} \,,
\label{eq:1p1hga-def}
\end{equation}
where 
\be
A^{\mu\nu}_{N} = \frac12 {\rm
Tr}\left[\left(\slashchar{p} + m_N \right)\gamma^0 \left( \Gamma^{\mu \rho}_N \right)^\dagger \gamma^0\left(\slashchar{p} +\slashchar{q} -\slashchar{k}_\gamma+ m_N\right) (\Gamma_N)^\nu_{.\,\rho} \right] \,,
\ee
in terms of the amputated $Z N \to \gamma N$ elementary amplitudes introduced in the previous section. This tensor is
evaluated at an average nucleon hole four momentum $p=\langle p \rangle$ to simplify the calculation. Our choice for $\langle p^\mu \rangle$ and discussions in support of this approximation are given in Refs.~\cite{Wang:2013wva,Hernandez:2013jka}. The definition and explicit expressions for the Lindhard function  $\overline {U}_R(q-k_\gamma,k_F^N,k_F^{N})$  can be found in Ref.~\cite{Nieves:2004wx}. It depends on the local density of protons or neutrons in the nucleus through the local Fermi momenta  $k^N_F(r)= [3\pi^2\rho_N(r)]^{1/3}$. 

For the coherent reaction, the nucleon wave functions are not altered by the interaction; the amplitudes should be summed over all nucleons in the target. This leads to~\cite{Wang:2013wva,Amaro:2008hd}
\be
W^{\mu\sigma}_{\rm coh} = -
  \frac{\delta(E_\gamma-q^0)}{64\pi^3M^2}  {\cal A}^{\mu\rho}
    \left({\cal A}^\sigma_{.\,\rho}\right)^*\,,
 \label{eq:zmunu}
\ee
with
\be
{\cal A}^{\mu\rho}(q,k_\gamma) = 
\int d^3r\ e^{i\left(\vec{q}-\vec{k}_\gamma\right)\cdot\vec{r}} 
\left\{ \rho_{p}(r) {\hat \Gamma}^{\mu\rho}_{p} + \rho_{n}(r\,)
 {\hat \Gamma}^{\mu\rho}_{n} \right\}
\label{eq:Jmunu2} \,,
\ee
and
\be
{\hat \Gamma}^{\mu\rho}_{N} = 
\frac12 {\rm Tr}\left[(\slashchar{p}+ m_N)\gamma^0\,\Gamma_{N}^{\mu\rho} \right] \left. \frac{m_N}{p^0}  
  \right|_{p^\mu=\left( \sqrt{m_N^2+\frac{(\vec{k}_\gamma-\vec{q}\,)^2}{4}},
      \frac12(\vec{k}_\gamma-\vec{q}\,)\right)}  \,, 
\label{eq:cn-jcoh} 
\ee  
in terms of the elementary $Z N \to N \gamma$ amputated amplitudes $\Gamma^{\mu\rho}_{N}$ of Eq.~(\ref{eq:hadronNR}) and Fig~\ref{fig:Feynman}. In these amplitudes, energy conservation  is accomplished by imposing
$q^0=E_\gamma$, which is justified by the large nucleus mass, while the transferred momentum is assumed to be equally shared between the initial and final nucleons. This approximation is discussed in Refs.~\cite{Alvarez-Ruso:2014bla,Wang:2013wva}. 

In both incoherent and coherent processes, the broadening of the $\Delta(1232)$ resonance
\be
\Gamma_\Delta \raw \tilde{\Gamma}_\Delta - 2\,  \mathrm{Im} \Sigma_\Delta(\rho)
\ee
in the nuclear medium is included. The resonance decay width is reduced to  $\tilde{\Gamma}_\Delta$ because the final nucleon in $\Delta \raw \pi N$ can be Pauli blocked but, on the other hand, is increased by many body processes $\Delta \, N \raw N \, N$, $\Delta \, N \raw N \,N \, \pi$ and  $\Delta \, N \, N \raw N \, N \, N$. For these decay channels, accounted in $\mathrm{Im} \Sigma_\Delta$, we have taken the parametrizations of Ref.~\cite{Oset:1987re}. This collisional broadening results in a reduction of the NC$\gamma$ cross sections in nuclei~\cite{Wang:2013wva}.

\subsection{Error budget}
\label{sec:errors}
  
The theoretical model described above has uncertainties in the treatment of both hadronic interactions and nuclear effects. We have performed an error analysis by propagating the uncertainties listed in Table~\ref{tab:err}, assuming that they are uncorrelated and Gaussian distributed. At the hadronic level, we consider errors in the leading $N-\Delta(1232)$ axial coupling $C^A_5(0)$ and in the parameter controlling its $q^2$ dependence (axial mass $M_{A\Delta}$) according to the analysis of weak pion production on deuterium performed in Ref.~\cite{Hernandez:2010bx}. The uncertainties in the $N-\Delta$ largest helicity amplitudes $A_{1/2}$ and $A_{3/2}$ at $q^2=0$ are also included by taking the relative errors from the PDG estimates~\cite{Beringer:1900zz}. The  unitary isobar model MAID~\cite{Drechsel:2007if}, from which our nucleon-to-resonance electromagnetic form factors are determined~\cite{Wang:2013wva}~\footnote{With the helicity amplitudes $A_{1/2}$ and $A_{3/2}$ at $q^2=0$ from MAID, one obtains a branching ratio of $\Gamma(\Delta \raw N\, \gamma)/\Gamma_{\mathrm{tot}}(\Delta) = 0.6 \%$, consistent with the PDG estimate of 0.55-0.65\%~\cite{Beringer:1900zz}.},
does not provide these errors. Uncertainties in the mechanisms  with $N^*$ intermediate states can be large, particularly in the axial transition currents which are poorly known. They can nevertheless be neglected because at the low energies of the peak in the T2K neutrino flux (Figure~\ref{fig:flux}), the contribution from these terms is quite small. Finally, in the case of the nucleon form factors present in the $NP$ and $CNP$ amplitudes, we neglect errors in the vector form factors and axial coupling but take into account the uncertainty in the  $q^2$ dependence of the axial form factor encoded in the axial mass $M_A$~\cite{Bodek:2007ym}. 

As explained above, our treatment of the reactions on nuclei relies on the local density approximation. For the distribution of protons in $^{16}$O we take the empirical harmonic oscillator parametrization, and the parameter errors, from Ref.~\cite{DeJager:1974dg}. For the distributions of neutrons, harmonic oscillator parametrizations and relative errors as in Ref.~\cite{DeJager:1974dg} are adopted  but with central values from Ref.~\cite{Nieves:1993ev}. As the in-medium $\Delta(1232)$ broadening plays an important role, we have gauged its uncertainty with a 10\% relative error. One should point out that our error estimate for nuclear effects does not account for possible multi-nucleon mechanisms that are not part of the model developed in Ref.~\cite{Wang:2013wva}.

\begin{table*}[htb!]
\caption{Error budget. \label{tab:err}}
\begin{center}
\begin{tabular}{c|c|c}
\hline\hline
Quantity &  Value & Source \\
\hline
$M_A$ & $1.016 \pm 0.026$~GeV & \cite{Bodek:2007ym} \\
$C^A_5(0)$ & $1.00 \pm 0.11$ & \cite{Hernandez:2010bx} \\
$M_{A\Delta}$ & $0.93 \pm 0.07$~GeV & \cite{Hernandez:2010bx} \\
$A_{1/2}$ & $(-140 \pm 6) 10^{-3}$~GeV$^{-1/2}$ & \cite{Drechsel:2007if,Beringer:1900zz} \\
$A_{3/2}$ & $(-265 \pm 5) 10^{-3}$~GeV$^{-1/2}$ & \cite{Drechsel:2007if,Beringer:1900zz} \\
$a_{p}$ & $1.833 \pm 0.014$~fm & \cite{DeJager:1974dg} \\
$\alpha_{p}$  & $1.544 \pm 0.001$~fm & \cite{DeJager:1974dg} \\
$a_{n}$ & $1.815 \pm 0.014$~fm & \cite{Nieves:1993ev,DeJager:1974dg} \\
$\alpha_{n}$  & $1.529 \pm 0.001$~fm & \cite{Nieves:1993ev,DeJager:1974dg} \\
$(\mathrm{Im} \Sigma_\Delta) r$  & $r = 1.0 \pm 0.1$ & \\
\hline\hline
\end{tabular}
\end{center} 
\end{table*}

\subsection{Photon events at SK}
\label{subsec:events}

\begin{figure}[h!]
\begin{center}
\includegraphics[width=0.45\textwidth]{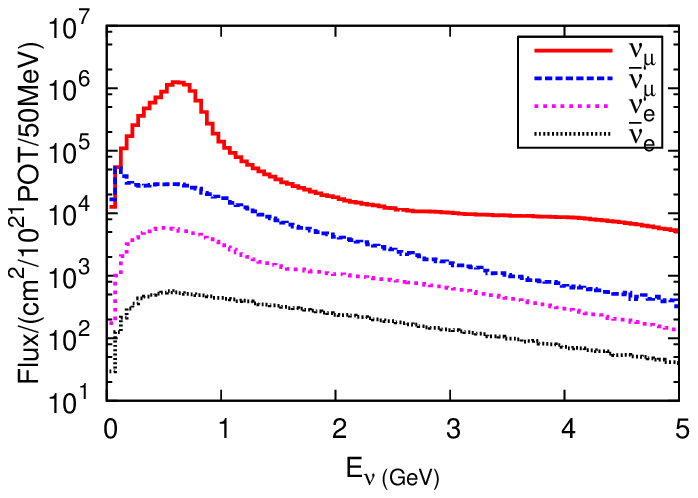}      
\caption{
\label{fig:flux}
(Color online) The T2K flux prediction at the SK detector~\cite{Abe:2012av} (without oscillations). It is shown only below 5 GeV although it is simulated up to 30 GeV.} 
\end{center}
\end{figure}

Once the differential cross sections for the components of the detector (H$_2$O) are established, it is straightforward to obtain the number of NC$\gamma$ events for a given photon energy and direction with respect to the neutrino beam, 
\be
\frac{dN}{dE_\gamma d\cos{\theta_\gamma}} = N_{\mathrm{POT}} \sum_{l=\nu,\bar\nu} \sum_{t=p,\,^{16}\mathrm{O}} N_t \int dE_\nu \phi_l(E_\nu) \frac{d\sigma_{l\, t}(E_\nu)}{dE_\gamma d\cos{\theta_\gamma}} \,.
\label{events}
\ee
Here, the total numbers of protons and $^{16}$O nuclei in the SK inner detector are 
\be
N_p = \frac{2}{18} M N_A = \frac{1}{9} M N_A \,, \qquad N_{^{16}\mathrm{0}} = \frac{16}{18} M \frac{N_A}{16} =\frac{1}{18} M N_A \,,
\label{Nt}
\ee 
where $M = 2.25  \times 10^{10}$ grams is the fiducial mass, and  $N_A$, the Avogadro number. Our estimate is for the recent T2K $\nu_e$ appearance analysis, corresponding to a total number of protons on target (POT) $N_{\mathrm{POT}} =  6.57 \times 10^{20}$ in $\nu$ mode~\cite{Abe:2013hdq}. The flux of the off-axis neutrino beam from Tokai has a narrow peak with median energy of 630 MeV at SK~\cite{Abe:2012av} (see Fig.~\ref{fig:flux}). We neglect its contribution above $E_\nu =3$~GeV. In spite of the rather long tail, a sizable contribution of the $E_\nu > 3$~GeV region would require a considerably large cross section at high energies, which we do not expect (see the discussion in Ref.~\cite{Alvarez-Ruso:2014bla}). The negative result in the single photon search performed by the NOMAD experiment~\cite{Kullenberg:2011rd}, with an average energy of the neutrino flux of $E_\nu \sim 25$~GeV, is in line with our assumption. As the NC interaction is flavor independent, the composition of the beam after oscillations can be ignored.

It will be instructive to consider also the (non-observable) neutrino-energy event distribution, which can be easily related to the integrated cross section
\be
\frac{dN}{dE_\nu} =  N_{\mathrm{POT}} \sum_{l=\nu,\bar\nu} \sum_{t=p,\,^{16}\mathrm{O}} N_t \phi_l(E_\nu) \sigma_{l\, t}(E_\nu) \,.
\label{events_int}
\ee

\section{Results}
\label{sec:result}

\begin{figure}[h!]
\begin{center}
\includegraphics[width=0.45\textwidth]{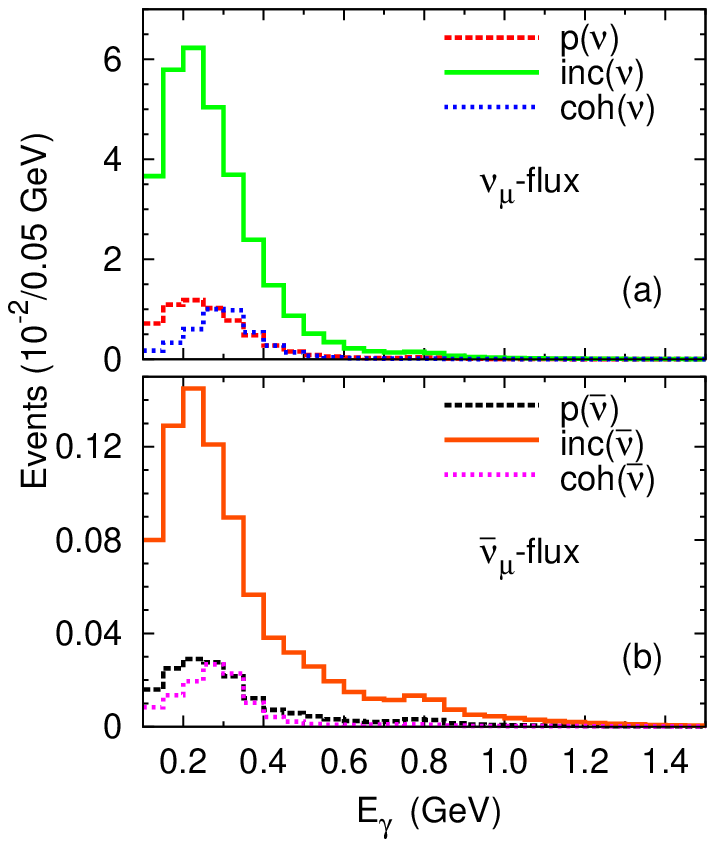}
\includegraphics[width=0.45\textwidth]{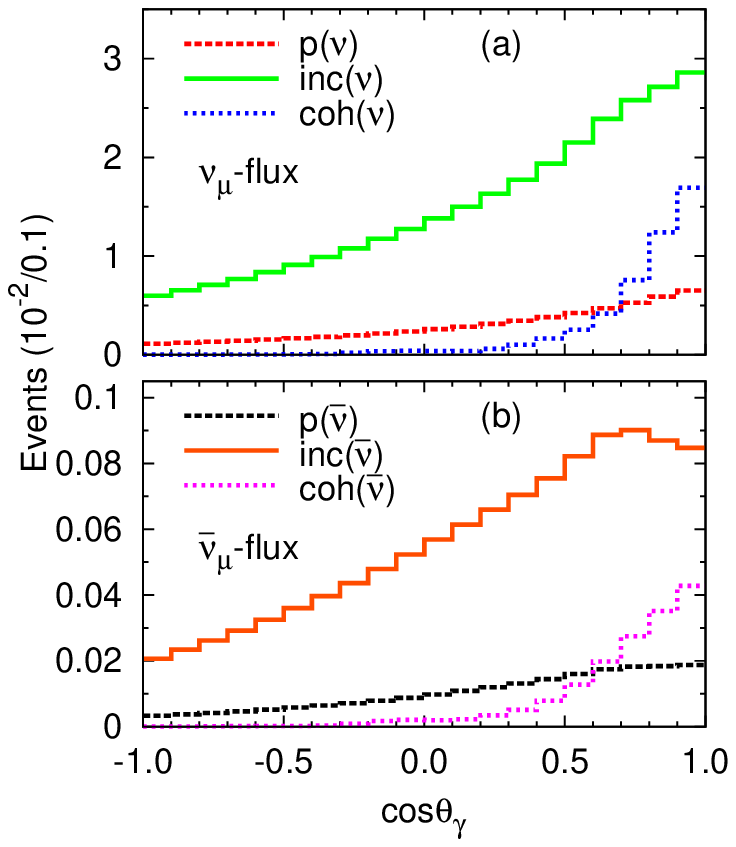}
\caption{(Color online) Photon energy (left) and angular (right) distributions of NC$\gamma$ events at SK from the T2K flux. The curves labeled as ``p'', ``inc'' and ``coh'' stand for the contributions of the $\nu(\bar \nu)-\rm{H}_2$, $\nu(\bar \nu)-^{16}\rm{O}$  incoherent and coherent reactions, respectively. Upper (lower) plots are for the  $\nu_\mu$ ($\bar \nu_\mu$) component of the flux.}
\label{fig:qf_th100}
\end{center} 
\end{figure}

The photon energy and angular distributions of the NC$\gamma$ events at SK are shown in Fig.~\ref{fig:qf_th100} for the different contributions, i.e. NC$\gamma$ on the two protons of H$_2$O and on $^{16}$O (coherent and incoherent). The contributions of the $\nu_\mu$ and $\bar\nu_\mu$ components of the flux are displayed.  The incoherent reaction is the largest and peaks at $E_\gamma \sim 200-300$~MeV, reflecting the importance of the $\Delta(1232)$. The yield of nucleons and the coherent channel, both similar in size, is smaller but still important. The  $\bar\nu_\mu$ yield is quite small while other flux components are totally negligible. 
The angular distributions are forward peaked, particularly for the coherent reaction. The latter has larger incidence than the nucleon channel in the forward direction. On the other hand, the angular dependence for the incoherent reaction induced by antineutrinos is softer than the neutrino one and peaks around $\cos\theta_\gamma = 0.7$. Similar features in the energy and angular distributions were obtained for the MiniBooNE detector~\cite{Wang:2014nat}.

Summing over all bins in the histograms above, one finds that the total number of NC$\gamma$ events is
\begin{equation}
\label{eq:total}  
{\cal N} = {\cal N}^{(\nu_\mu)} +  {\cal N}^{(\bar\nu_\mu)} = (0.412 \pm 0.049) + (0.015 \pm 0.002) = 0.427 \pm 0.050  \,.
\end{equation}
The errors correspond to a standard 68\% confidence level (CL) and are dominated by the uncertainty in  $C^A_5(0)$. This is a small quantity compared to the 28 e-like events detected at SK~\cite{Abe:2013hdq}~\footnote{The comparison is only indicative due to the lack of efficiency correction in our estimate.} but can be relevant in future attempts to measure $\delta_{CP}$. This result becomes even more significant when compared to the NEUT~5.1.4.2 equivalent figure of
\begin{equation}
{\cal N}_{\rm NEUT} =  {\cal N}_{\rm NEUT} ^{(\nu_\mu)} +  {\cal N}_{\rm NEUT} ^{(\bar\nu_\mu)}  = 0.209 + 0.008  = 0.217 \,.
\end{equation}
Indeed, using the NC$\gamma$ cross section model of Ref.~\cite{Wang:2013wva} we predict nearly twice more events than NEUT~5.1.4.2, on which the T2K estimate is based. In NEUT, NC$\gamma$ interactions proceed via baryon resonance excitation (predominantly $\Delta(1232)$) followed by radiative decay. Resonance production is implemented according to the Rein-Sehgal original model~\cite{Rein:1980wg} but with the parameter $m_A =1.21$~GeV in the transition form factors of Eq.~(3.12) in that reference. 

\begin{figure}[h!]
\begin{center}
  \includegraphics[width=0.45\textwidth]{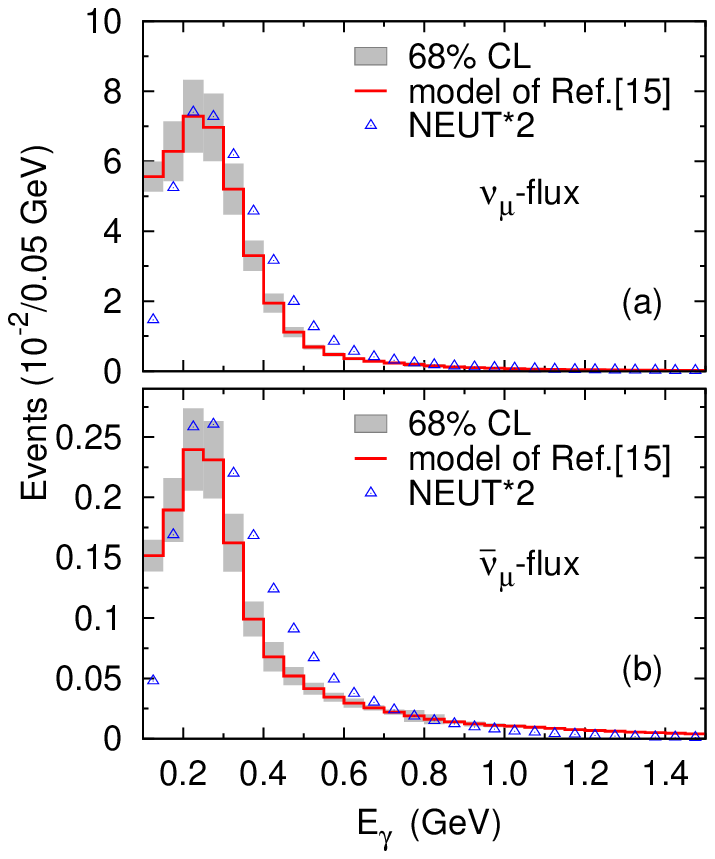}
\includegraphics[width=0.45\textwidth]{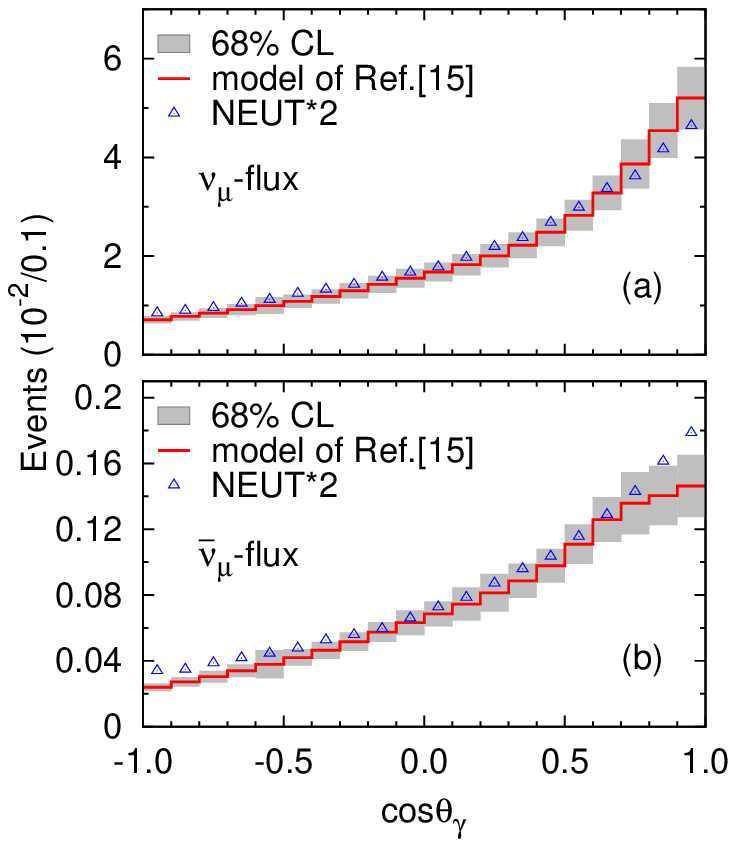}
\caption{(Color online) $E_\gamma$ and $\cos{\theta_\gamma}$ distributions of NC$\gamma$ events at SK for the  T2K experiment, obtained with the model of Ref.~\cite{Wang:2013wva} (solid line) and with NEUT~5.1.4.2~\cite{Hayato:2009zz} (triangles).  The bands correspond to a 68\% CL based on the error budget of Sec.~\ref{sec:errors}. For a better shape comparison, NEUT results have been multiplied by a factor of 2.}
\label{fig:qfth}
\end{center}
\end{figure}

\begin{figure}[h!]
\begin{center}
  \includegraphics[width=0.45\textwidth]{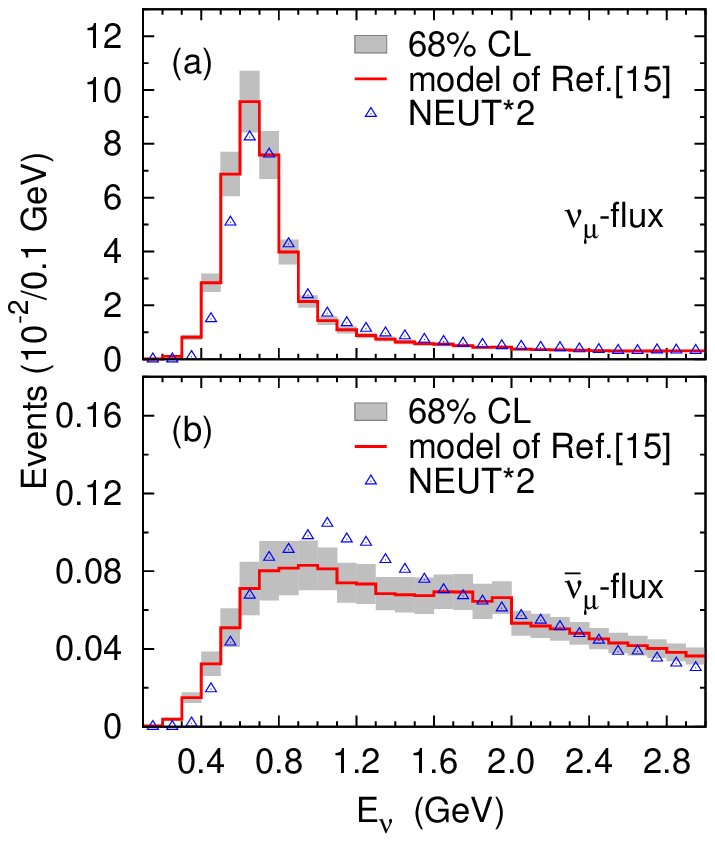} 
  \caption{(Color online) $E_\nu$ distribution of NC$\gamma$ events at SK for the T2K experiment. Line styles have the same meaning as in Fig.~\ref{fig:qfth}. The results from the NEUT generator have been multiplied by a factor of 2.}
\label{fig:enu}
\end{center}
\end{figure}

In view of this discrepancy, we have performed more detailed comparisons with the NEUT~5.1.4.2 NC$\gamma$ prediction and confronted photon energy and angular distributions. Figure~\ref{fig:qfth} does not reveal significant shape differences between the two models although the one of Ref.~\cite{Wang:2013wva} predicts considerably more events at $E_\gamma = 0.1-0.2$~GeV and a faster decrease for $E_\gamma \approx 0.3-0.6$~GeV. For the $\bar\nu_\mu$ flux, the photon angular distribution from the model of  Ref.~\cite{Wang:2013wva} is slightly flatter than the one from NEUT.
The neutrino energy dependence of the events, displayed in Fig.~\ref{fig:enu}
also shows a good agreement in the shape although for the $\bar\nu_\mu$ flux, the results by the model of Ref.~\cite{Wang:2013wva} do not show a peak around $E_{\bar\nu_\mu} \sim 1$~GeV. In both models, most events come from neutrinos with energies between 0.4 and 1.2~GeV where the approach, choice of mechanisms and approximations of Ref.~\cite{Wang:2013wva} are applicable. From the $E_{\bar\nu_\mu}$ distributions, it is apparent that the yield from $\bar\nu_\mu$ is underestimated when only $E_{\bar\nu_\mu} < 3$~GeV are kept because a sizable fraction of the $\bar\nu_\mu$ flux is left out. Nevertheless, this does not affect the comparisons, that have been consistently performed, nor the predictions for the total number of events, which are largely dominated by the $\nu_\mu$ contribution. Indeed,  ${\cal N}^{(\bar\nu_\mu)}$ in Eq.~(\ref{eq:total}) is smaller than the error in ${\cal N}$.

The comparisons in Figs.~\ref{fig:qfth} and \ref{fig:enu} indicate that the disagreement is due to a discrepancy in the size (normalization) of the integrated cross sections in the two models. This is consistent with the comparison of the NC$\gamma$ integrated cross sections on $^{12}$C from different models, as a function of $E_\nu$ displayed in Fig.~9 of Ref.~\cite{Katori:2014qta}.

\section{Conclusions}
\label{sec:conclusions}

The microscopic model of Ref.~\cite{Wang:2013wva} for single photon emission in (anti)neutrino NC interactions has been applied to predict the number of such events at the inner SK water Cherenkov detector, as well as their energy and angular distributions. With this  model one can take into account not only the radiative decay of weakly excited $\Delta(1232)$ resonance on both nucleons and nuclei, but also smaller, although relevant, contributions from nucleon pole terms and the coherent channel (details can be found in Ref.~\cite{Wang:2013wva}). For a $N_{\mathrm{POT}} =  6.57 \times 10^{20}$ we predict $0.427 \pm 0.050$ events without efficiency corrections; only 3\% of it arises from the $\bar\nu_\mu$ contamination of the $\nu_\mu$ beam. This small but irreducible background should be realistically estimated in order to increase the precision in the determination of oscillation parameters, particularly in $\delta_{CP}$ measurements. 

Remarkably, the prediction based on the model of Ref.~\cite{Wang:2013wva} is twice larger than the one obtained from the main T2K Monte Carlo generator NEUT~\cite{Hayato:2009zz} (version 5.1.4.2). In a detailed inspection, we have found no significant differences in the shapes of the photon energy, photon angular and neutrino energy distributions in the two models. The large difference in normalization cannot be solely attributed to the lack of non-$\Delta$ production amplitudes or coherent photon emission in NEUT. The same is true for the different treatment of nuclear corrections. A closer inspection of the dominant $\Delta(1232)$ mediated mechanism on the nucleon indicates that the largest mismatch arises, not from the $\Delta(1232)$ production, common to weak pion production reactions, but from the strength of the $\Delta(1232)$ radiative decay, which is smaller in NEUT~5.1.4.2. It is also worth recalling that, as shown in Ref.~\cite{Wang:2014nat}, the number of NC$\gamma$ events at the MiniBooNE detector predicted by the same microscopic model used here is consistent with the MiniBooNE {\it in situ} estimate obtained with the NUANCE generator~\cite{Casper:2002sd} tuned to the NC$\pi^0$ measurement~\cite{AguilarArevalo:2009ww}.

\section*{Acknowledgments}
\label{sec:acknow}

We thank H. A. Tanaka and S. Tobayama for useful discussions and support. One of us, LAR, thanks S. E. A. Orrigo for her assistance with ROOT. 
The work of LAR has been partially supported by the U.S. Department of Energy, Office of Science, Office of High Energy Physics, through the Fermilab Intensity Frontier Fellows Program. He gratefully acknowledges the hospitality during his stay at Fermilab.
This research has been partly supported by the Japanese Ministry of Education, Culture, Sports, Science and Technology, the 
Spanish Ministerio de Econom\'ia y Competitividad and European FEDER funds, under contracts FIS2011-28853-C02-02 and FIS2014-51948-C2-1-P, Generalitat Valenciana under program Prometeo II, 2014/068 and the Spanish Excellence Network on Hadronic Physics, FIS2014-57026-REDT. We also acknowledge support from the European Union project Study of Strongly Interacting Matter (acronym HadronPhysics3, Grant Agreement 283286) under the Seventh Framework Program.

\bibliography{neutrinos}

\end{document}